%% file: DMequi.tex
\documentclass[showpacs,twocolumn,prb,floatfix]{revtex4-1}

\usepackage[utf8]{inputenc}
\usepackage[T1]{fontenc}

\usepackage{graphicx, color, rotating} 
\usepackage{dcolumn} 

\usepackage{tikz}
\pgfrealjobname{DMequi}
\usetikzlibrary{automata,arrows,calc,decorations,decorations.pathreplacing,decorations.markings,chains,positioning,fit,backgrounds,fadings,patterns,matrix,shapes,shapes.symbols}
\usepackage{pgfplots}
\input{Abbildungen/tikzstyles}

\usepackage{colorthemekit}

\usepackage{amsmath}
\usepackage{amssymb}
\usepackage{dsfont}
\usepackage{mathtools}
\usepackage{nicefrac}
\usepackage{braket}
\usepackage{trfsigns} 
\usepackage[scr=boondoxo]{mathalfa} 

\usepackage[textsize=tiny]{todonotes}

\usepackage[plainpages=false, pdfpagelabels, bookmarksopen, pdfborder={0 0 0}, linkcolor=LinkColor, linkbordercolor={0 0 1},breaklinks]{hyperref}

\definecolor{LinkColor}{rgb}{0.00,0.00,0.00}
\usepackage[all]{hypcap} 
\DeclareMathOperator{\Tr}{Tr}
\DeclareMathOperator{\adj}{adj}
\DeclareMathOperator{\Res}{Res}

\begin{document}

\title{Distortion of a reduced equilibrium density matrix: influence on quantum emulation}

\author{Iris Schwenk and Michael Marthaler}
\affiliation{Institut f\"ur Theoretische Festk\"orperphysik, Karlsruhe Institute of Technology, D-76128 Karlsruhe, Germany}


\date{\today}

\begin{abstract}
We study a system coupled to external degrees of freedom, called bath, where we assume that the total system consisting of system and bath is in equilibrium.
An expansion in the coupling between system and bath leads to a general form of the reduced density matrix of the system as a function of the bath self energy.
The coupling to the bath results in a renormalization of the energies of the system and in a change of the eigenbasis.
This theory is applicable to quantum emulators  in thermal equilibrium.
Undesired external degrees of freedom can affect their reliability. We study the influence of bosonic degrees of freedom on the state of a six qubit system. 
\end{abstract}

\maketitle 

\section{Introduction} 

The research on quantum computation has lead to a tremendous development of controlled quantum systems\cite{Brumfiel2012}.
Even though building a universal quantum computer remains a huge challenge\cite{Fowler2012}, it is now possible to build artificial quantum systems to explore special physical models.
This method is called quantum emulation or analog quantum simulation\cite{Manousakis2002a}.
It differs from digital quantum simulation, where the time evolution of a system is simulated using qubit gates.
A comprehensive review of quantum simulation is given in Ref. \onlinecite{Georgescu2014a}.
In quantum emulation the idea is to map a Hamiltonian on an artificial quantum system, which can be controlled and read out.
One simple example is to map spin problems to coupled two level systems\cite{Britton2012}, which could be build by e.g. using trapped ions or superconducting qubits\cite{Macha2014}.
These systems should be easier to build than a universal quantum computer but provide the possibility to solve problems which can not be solved numerically on classical computers.

Superconducting circuits allow for a vast flexibility and several quantum emulation proposals exist\cite{Chen2014}.
But especially in the field of cold gases a substantial amount of spectacular experiments have been performed.
The problems studied range from strongly correlated fermionic systems\cite{Ohara2002} over nonequilibrium physics\cite{Cui2015} to strongly correlated bosonic systems\cite{Endres2012,Li2015} and frustrated magnetism\cite{Struck2011}.
Already such experiments reach the limit of classical computation\cite{Hart2014}, in the sense that present-day classical computers cannot perform first-principles simulations of these large systems.
But quantum emulation cannot just be used for condensed matter problems.
There is even the possibility to solve problems from high energy physics or cosmology\cite{Alsing2005,Giovanazzi2005,Nation2009,Lee2015a}.

In order to use quantum emulation to solve new problems, we need to estimate the reliability of emulation experiments.
There are two approaches\cite{Hauke2012} to validate results of a quantum emulation.
One is called cross validation\cite{Leibfried2010}.
This means that the emulation is performed on different physical realizations.
One expects consistent results for universal properties, because it is very likely that the errors are not the same in different realizations.
The second method is to use analytically or numerically obtained results to validate the result of the emulation\cite{Trotzky2009,Jordens2010}.
However this requires the emulation to be performed in a regime where analytical or numerical calculation is possible, defeating the purpose of quantum emulation. 
The hope is of course that after validation of the results of a quantum emulation in a classically solvable regime, the emulator works also in other regimes.
But both validation techniques give no quantitative evaluation of the error, and they restrict the usability of quantum emulation.

If a quantum emulator, described by the ideal Hamiltonian $H_S$, is studied in equilibrium conditions, we know that the ideal form of the density matrix of the emulator is given by $\rho_S=e^{-\beta H_S}/{{\Tr}(e^{-\beta H_S}) }$.
However if the coupling of the emulator to external degrees of freedom is not infinitesimally small, the form of the equilibrium density matrix can be changed\cite{Zurek1981,Review1982}.
Also, in the context of adiabatic quantum computation the influence of external degrees of freedom has been studied\cite{Sarandy2005a, Ashhab2006, Saito2007}.
For classical and quantum systems similar approaches to ours have already been used to analyze the reduced density matrix of a system with finite coupling to its environment \cite{Deng2013,Jarzynski2004}.
We analyze a system $H_S$ which is the ideal quantum emulator, coupled to external degrees of freedom like phonons or the electromagnetic field.
The full Hamiltonian consists of three parts,
\begin{align}
H=H_S+H_C+H_B
\end{align}
where the external degrees of freedom are contained in the bath Hamiltonian $H_B$, and a coupling between bath and ideal emulator is described by $H_C$.

The eigenstates and eigenenergies of the system are given by
\begin{align}
H_S\ket{s}=E_s\ket{s}\, .
\end{align}
In this paper we regard this setting in thermal equilibrium.
Therefore we know that the form of the density matrix is given by $\rho=e^{-\beta H}/{{\Tr}(e^{-\beta H})}$.
If it is our goal to determine an equilibrium expectation value for an operator $A$ acting on the emulator described by $H_S$,
we measure
\begin{align}
\braket{A} &= \Tr\left( \rho A \right)=\Tr_S\left( \tilde{\rho}_S A \right)
\end{align}
with $\tilde{\rho}_S ={\Tr}_B(\rho)$, and $\Tr_{S}$ ($\Tr_B$) is the trace over the states of the emulator (bath).
In general this expectation value deviates from the unperturbed value $\Tr_S\left( \rho_S A \right)$ because $\tilde{\rho}_S \neq \rho_S$.
To understand the influence of the external degrees of freedom on measurement results we therefore have to calculate the reduced density matrix $\tilde{\rho}_S$.

\smallskip
Using a diagrammatic technique we find the following form for the components of the reduced density matrix
\begin{align}
\braket{s|\tilde{\rho}_S |s'}=\sum_n f^{ss'}_n(\beta)e^{-\beta E_n^{re}} \, , \label{eq:fss'}
\end{align}
where $E_n^{re}$ denotes the renormalized energy.
This means that the coupling to external degrees of freedom has two consequences.
On the one hand it leads to a renormalization of energies, where the renormalized energy depends on diagonal elements of the self energy $\Sigma$ which describes the influence of external degrees of freedom.
\begin{align}
E_s^{re} = E_s-\Sigma_{ss}(-E_s)
\end{align}
On the other hand the eigenbasis of the density matrix changes since $f^{ss'}_n$ is in general not zero for $s \neq s'$.
A detailed discussion of Eq.~(\ref{eq:fss'}) is given in Sec.~\ref{sec:general_solution}.

\smallskip
In Sec.~\ref{sec:diagrammatic_expansion} we describe the diagrammatic expansion leading to an equation for the reduced density matrix.
The solution of this equation using the Laplace transform is presented in Sec.~\ref{sec:general_solution}.
Finally in Sec.~\ref{sec:lowest_order_results} we focus on the results for the lowest order in self energy and consider a six qubit model system.

\section{Diagrammatic Expansion} \label{sec:diagrammatic_expansion}

The equilibrium state of the full system is described by the density matrix,
\begin{align}
\rho=\frac{1}{Z} e^{-\beta H}
\end{align}
where $Z$ denotes the corresponding partition function.
The state of the system under the influence of the environment is given by the reduced density matrix, where the trace is taken over the external degrees of freedom,
\begin{align}
\tilde{\rho}_S = \Tr_B\left( \rho \right) \neq \frac{1}{Z_S} e^{-\beta H_S}  = \rho_S \; .
\end{align}
In general the reduced density matrix $\tilde{\rho}_S$ differs from the equilibrium density matrix of the system Hamiltonian $\rho_S$. This discrepancy describes the influence of the external degrees of freedom.

Using the interaction picture in imaginary time
\begin{align}
H_C(\tau) = e^{\tau(H_S+H_B)} \, H_C \, e^{-\tau(H_S+H_B)}
\end{align}
we find the following identity which is useful to rewrite the equation for the reduced density matrix,
\begin{align}
e^{-\beta H}=e^{-\beta (H_S+H_B)}&\underbrace{\mathcal{T}\exp{\left(-\int\limits_{0}^{\beta}\mathrm{d}\tau \; H_C(\tau)\right)}}_{=\ \mathcal{S}} \label{eq:Htilde}
\end{align}
where $\mathcal{T}$ denotes the time ordering operator.
A similar approach based on this equation was used in the context of adiabatic quantum computation by Deng et al. in Ref. \onlinecite{Deng2013}.
But they only calculated the lowest order term in $H_C$, while we formulate our expansion to all orders. 
As we will show later, we can sum certain classes of diagrams to all orders. 

Using Eq.~(\ref{eq:Htilde}) we obtain the reduced density matrix $\tilde{\rho}_S$ expressed in terms of the equilibrium density matrix of the system Hamiltonian $\rho_S$ and the bath Hamiltonian $\rho_B$,
\begin{align} 
\tilde{\rho}_S &= \Tr_B\left( \frac{1}{Z} e^{-\beta H} \right)\nonumber\\
&=\Tr_B\left(\frac{1}{Z} e^{-\beta (H_S+H_B)} \cdot \mathcal{S} \right)\nonumber\\
&= \rho_S \, \frac{Z_S Z_B}{Z} \, \Tr_B\left( \rho_B \mathcal{S}\right) \;.
\end{align}
Thus it appears that $\tilde{\rho}_S$ depends on the unperturbed density matrix $\rho_S$ and a bath expectation value which reflects the effect of the coupling $H_C$.
Using Eq.~(\ref{eq:Htilde}) the fraction of partition functions can be identified as the so called vacuum diagrams $\braket{\mathcal{S}}_o$, where $\braket{\dots}_o = \Tr{(\rho_S\rho_B\dots)}$,
\begin{align}
\braket{\mathcal{S}}_o = \Tr\left( \rho_S\rho_B \mathcal{S} \right) = \frac{Z}{Z_S Z_B} \;.
\end{align}
This allows us to rewrite the reduced density matrix in terms of $\rho_S$ and $\mathcal{S}$,
\begin{align}
\tilde{\rho}_S = \frac{1}{\braket{\mathcal{S}}_o} \, \rho_S \, \braket{\mathcal{S}}_B \; . \label{eq:rhoende}
\end{align}
An expansion of $\mathcal{S}$ in the coupling Hamiltonian can be used to find an approximation for the reduced density matrix $\tilde{\rho}_S$,
\begin{align}
\mathcal{S} \approx 1 - \underbrace{\int\limits_0^\beta H_C(\tau) \mathrm{d}\tau}_{Tr_B \Rightarrow 0} +\int\limits_0^\beta \mathrm{d}\tau_1 \int\limits_0^{\tau_1} \mathrm{d}\tau_2 H_C(\tau_1) H_C(\tau_2) + \dots \; .
\end{align}
We can always set $\braket{H_C}_B$ to zero, by adding an appropriate term to the system Hamiltonian. Therefore odd powers of $H_C$ vanish in the expectation value.
To perform this expansion we use a diagrammatic technique.
We assume the coupling Hamiltonian to be of the form,
\begin{align}
H_C = \sum_j \mathscr{O} \cdot t_j (b_j^\dagger + b_j) = \sum_j \mathscr{O} \cdot q_j \label{eq:HC}
\end{align}
where $\mathscr{O}$ is an operator acting on the system, and $b_j^{\dagger},b_j$ denote creation and annihilation operators of bosonic modes with frequency $\omega_j$.
Modeling the bath as a sum of bosonic modes, allows for the description of every bath where fluctuations have a Gaussian distribution and in most cases non-Gaussian corrections remain small \cite{Schad2014,Golubev2010}.
For superconducting qubits the bath could also be fermionic \cite{Catelani2011,Leppakangas2011}, which could also be described by our theory with only a few adjustments. 
If for all contractions only equivalent pairs of annihilation and creation operators are chosen \cite{Haenggi_Schoen_Tunneling}, 
the only change is that the spectral function of the bath is given by  Eq. (20) in Ref. \onlinecite{Zanker2015}. For a more general approach contraction rules
have to be defined with an appropriate sign change \cite{Schoeller_Tunneling}.

We obtain the diagrammatic rules by examining the expansion of one matrix element of the reduced density matrix,
\begin{align}
\label{eq:entwicklung}
\braket{s'|\tilde\rho_S|s} &\approx \frac{1}{Z_S \braket{\mathcal{S}}_o} e^{-(\beta-0)E_{s'}} \ \delta_{s,s'} \nonumber\\+&  \frac{1}{Z_S \braket{\mathcal{S}}_o} \int\limits_0^\beta \mathrm{d}\tau_1 \int\limits_0^{\tau_1}\mathrm{d}\tau_2 \; \sum\limits_{\bar{s},j} e^{-(\beta-\tau_1)E_{s'}} \braket{s'|\mathscr{O}|\bar{s}} \nonumber\\ &\cdot e^{-(\tau_1-\tau_2)E_{\bar{s}}} \braket{\bar{s}|\mathscr{O}|s} e^{-(\tau_2-0)E_s} \braket{q_j(\tau_1)q_j(\tau_2)}_B \nonumber\\+&\mathcal{O}(H_C^4) \; .
\end{align}
Here we have assumed, that different bath modes are not correlated,
\begin{align}
\braket{q_j(\tau_1)q_k(\tau_2)}=0 \quad \text{for} \quad j\neq k \; .
\end{align}
Free propagations $e^{-(\tau_{left}-\tau_{right})H_S}$ are represented by straight lines.
A dashed line means a bath correlator $\sum_j\braket{q_j(\tau_{left})q_j(\tau_{right})}_B$.
At the beginning and the end of a dashed line we have an operator $\mathscr{O}$.
The order of all operators is denoted by the diagram.
At every full vertex there is an integration $\int_0^{\tau_{n-1}}\mathrm{d}\tau_n$, whereas $\tau_{n-1}$ denotes the time left of $\tau_n$.
Every diagram starts at the right with $\tau=0$ and ends at the left with $\beta$.
If a diagram does not start at $\tau=0$ we have to change the borders of integration to obey time ordering.
The diagrammatic series representing the reduced density matrix $\tilde{\rho}_S^{(u)}$ (Eq.~(\ref{eq:entwicklung})), which we obtain using this rules is the following,
\begin{align}
\begin{tikzpicture}[baseline=(current bounding box.center)]
\input{Abbildungen/diagramme.tex}
\end{tikzpicture}
\end{align}
For simplicity we introduced the unnormalized reduced density matrix $\tilde{\rho}_S^{(u)}= \tikz[baseline=0.41cm]{\input{Abbildungen/vollesrho.tex}}$,
\begin{align}
\tilde{\rho}_S^{(u)}=Z_S\braket{S}_o \tilde{\rho}_S \; .
\end{align}
We can define a self energy,
\begin{align}
\begin{tikzpicture}[baseline=(current bounding box.center)+5pt]
\input{Abbildungen/selbstenergie.tex}
\end{tikzpicture}
\end{align}
 and obtain a Dyson equation,
\begin{align}
\begin{tikzpicture}[baseline={([yshift=-0.5pt]current bounding box.center)}]
\input{Abbildungen/dyson.tex}
\end{tikzpicture}
\; .
\end{align}
The explicit expression for the Dyson equation is given by
\begin{align}
&\tilde{\rho}_S^{(u)}(\beta) \nonumber\\  &= e^{-\beta H_S} + \int\limits_0^\beta \mathrm{d}\tau_1 \int\limits_0^{\tau_1} \mathrm{d}\tau_2 e^{-(\beta-\tau_1)H_S} \Sigma(\tau_1-\tau_2) \tilde{\rho}_S^{(u)}(\tau_2) \; . \label{eq:dysoneq}
\end{align}
After solving this equation a normalization according to 
\begin{align}
\Tr_S\tilde{\rho}_S = 1
\end{align}
gives us the reduced density matrix $\tilde{\rho}_S$.
A derivation with respect to $\beta$ leads to an equation of motion for the reduced density matrix,
\begin{align}
\frac{\partial}{\partial \beta} \tilde{\rho}_S^{(u)}(\beta) = - H_S \tilde{\rho}_S^{(u)}(\beta) + \int\limits_0^\beta \mathrm{d}\tau_2 \,\Sigma(\beta-\tau_2)\tilde{\rho}_S^{(u)}(\tau_2) \label{eq:maineq}
\end{align}
which is quite similar to the well-known master equation. However the inverse temperature $\beta$ takes the role of time and the self energy is simply an operator and not a superoperator.

The explicit expression for the lowest order term of the self energy is given as follows,
\begin{align}
\Sigma^{(1)}(\tau_{l}-\tau_{r})=& \sum_{\bar{s},j}\mathscr{O}\ket{\bar{s}}e^{-(\tau_{l}-\tau_{r})E_{\bar{s}}}\bra{\bar{s}}\mathscr{O} \braket{q_j(\tau_{l})q_j(\tau_{r})}_B \, . \label{eq:self energywithcorr}
\end{align}
Defining the spectral density of the bath modes
\begin{align}
\sum_j\braket{q_j(\tau_1)q_j(\tau_2)}_B = \int \frac{\mathrm{d}\omega}{2\pi} S(\omega) e^{-\omega(\tau_1-\tau_2)} \; ,\label{eq:correlator}
\end{align}
we can rewrite the self energy, which takes the form
\begin{align}
&\Sigma^{(1)}(\tau_{l}-\tau_{r})\nonumber\\ &= \int \frac{\mathrm{d}\omega}{2\pi} \sum_{\bar{s}}\mathscr{O}\ket{\bar{s}}e^{-(\tau_{l}-\tau_{r})E_{\bar{s}}}\bra{\bar{s}}\mathscr{O}\; S(\omega) e^{-\omega(\tau_{l}-\tau_{r})} \; .
\end{align}

\subsection{Truncation of the series for $\tilde{\rho}^{(u)}_S$}
\begin{figure}[t]
\centering
\includegraphics{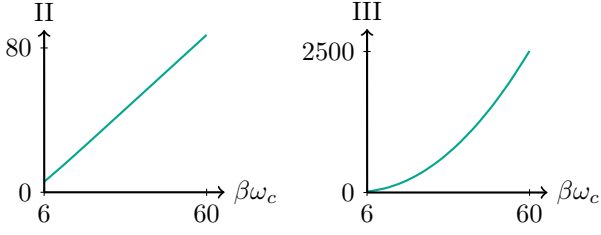}
\caption{The graphs show the $\beta$ dependence of diagrams II and III.}
\label{fig:plot_abbruch_reihe}
\end{figure}

To understand the conditions for the convergence of our expansion we consider diagrams of first and second order in $H_C$.
\begin{align}
\begin{tikzpicture}[baseline={([yshift=-18pt]current bounding box.center)}]
\input{Abbildungen/abbruch_reihe.tex}
\end{tikzpicture}
\; .
\end{align}
As a worst case estimation we set the contributions of the free propagations and the matrix elements of $\mathscr{O}$ to unity.
We focus on the limit $\beta\rightarrow\infty$.
In this regime the diagrams are dominated by the following terms,
\begin{align}
\text{I} &\propto \int \frac{\mathrm{d}\omega_1}{2\pi} \; \beta \frac{S(\omega_1)}{\omega_1}  \\
\text{II} &\propto \iint \frac{\mathrm{d}\omega_1}{2\pi}  \frac{\mathrm{d}\omega_2}{2\pi} \; \beta \frac{S(\omega_1) S(\omega_2)}{\omega_1^2 \omega_2 + \omega_1 \omega_2^2}  \\
\text{III} &\propto \iint \frac{\mathrm{d}\omega_1}{2\pi}  \frac{\mathrm{d}\omega_2}{2\pi} \;\beta^2 \frac{S(\omega_1) S(\omega_2)}{2\omega_1 \omega_2} \; .
\end{align}
For the spectral density, we assume an ohmic spectrum with Lorentzian cutoff at frequency $\omega_c$,
\begin{align}
S(\omega) = \frac{\eta \omega}{(1-e^{-\beta\omega})(1+(\omega/\omega_c)^2)} \label{eq:spectraldensity}
\end{align}
whereas $\eta$ denotes the coupling strength.
We introduce dimensionless parameters $\nu_{1/2} = \nicefrac{\omega}{\omega_c}$ and $\tilde{S}(\nu) = \nicefrac{S(\nu)}{\eta\omega_c}$ to find
\begin{align}
\text{I} &\propto \eta\beta\omega_c \int \frac{\mathrm{d}\nu_1}{2\pi} \; \beta \frac{\tilde{S}(\nu_1)}{\nu_1}  \\
\text{II} &\propto \eta^2\beta\omega_c \iint \frac{\mathrm{d}\nu_1}{2\pi}  \frac{\mathrm{d}\nu_2}{2\pi} \; \frac{\tilde{S}(\nu_1)\tilde{S}(\nu_2)}{\nu_1^2 \nu_2 + \nu_1 \nu_2^2}  \\
\text{III} &\propto \eta^2\beta^2\omega_c^2 \iint \frac{\mathrm{d}\nu_1}{2\pi}  \frac{\mathrm{d}\nu_2}{2\pi} \; \frac{\tilde{S}(\nu_1)\tilde{S}(\nu_2)}{2\nu_1 \nu_2} \; .
\end{align}
To verify this behavior, Fig.~\ref{fig:plot_abbruch_reihe} shows a numerical calculation for diagrams II and III, where we consider all orders in $\beta$.
We immediately see, that diagram II can be neglected in comparison to I if $\eta\ll1$.
However diagram III is only negligible in comparison to I if
\begin{align}
\eta\beta\omega_c \ll 1 \; .
\end{align}
Considering this we see that it is necessary to sum up separable diagrams within the self energy, especially in the limit $\beta\rightarrow\infty$, and that it is possible to neglect fully crossed diagrams like I.
Because of the similarity of the self energy to the diagrams discussed above
\begin{align}
\begin{tikzpicture}[baseline=(current bounding box.center)+5pt]
\input{Abbildungen/selbstenergie_abbruch.tex}
\end{tikzpicture}
\; ,
\end{align}
it is possible to truncate the series for the self energy if $\eta < 1$.

\subsection{Generalization of $H_C$} \label{sec:generalization}
To derive the diagrammatic theory, we assumed that the coupling Hamiltonian has a simple form (Eq.~(\ref{eq:HC})).
However, in the case of a system consisting of many qubits, it is reasonable to assume that every qubit is coupled to an individual bath.
Therefore we consider a generalized coupling Hamiltonian,
\begin{align}
H_C = \sum_{n,j} \mathscr{O}^{(n)} q_j^{(n)} = \sum_n h_n \; .
\end{align}
We assume all baths to be of the same type,
\begin{align}
\braket{q_j^{(n)}(\tau_1)q_j^{(n)}(\tau_2)}_B = \braket{q_j^{(k)}(\tau_1)q_j^{(k)}(\tau_2)}_B \; .
\end{align}
But the individual baths are not correlated,
\begin{align}
\braket{q_j^{(n)}(\tau_1)q_j^{(k)}(\tau_2)}_B = 0 \quad\text{for} \quad n\neq k \; .
\end{align}
In this case the self energy is given as a sum of self energies for each coupling operator $h_n$,
\begin{align}
\Sigma = \sum_n \Sigma_{n} \; .
\end{align}

\section{General solution} \label{sec:general_solution}

Due to the convolution integral in Eq.~(\ref{eq:maineq}), a solution in Laplace space is appropriate,
\begin{align}
\tilde{\rho}_S^{(u)}(\epsilon) = \int\limits_0^\infty \tilde{\rho}_S^{(u)}(\tau) \, e^{-\epsilon\tau} \, \mathrm{d}\tau
\; \text{, with } \; \epsilon=i\omega+\eta \; .\label{eq:trafo}
\end{align}
The equation for the unnormalized reduced density matrix in Laplace space reads
\begin{align}
\epsilon \tilde{\rho}_S^{(u)}(\epsilon) -\tilde{\rho}_S^{(u)}(\tau=0) = \left(- H_S + \Sigma(\epsilon) \right)\tilde{\rho}_S^{(u)}(\epsilon) \; .\label{eq:laplacmaineq}
\end{align}
This directly leads to the solution in Laplace space,
\begin{align}
\tilde{\rho}_S^{(u)}(\epsilon) = \left( \epsilon + H_S - \Sigma(\epsilon) \right)^{-1} \tilde{\rho}_S^{(u)}(\tau=0) \;  \label{eq:solutionlaplace}.
\end{align}
The imaginary time corresponds to the temperature. Therefore $\tau=0$ means an infinitely high temperature, which means in turn equal probabilities for each state. Hence we assume $\tilde{\rho}_S^{(u)}(\tau=0) \propto \mathds{1}$.

To get an hermitian density matrix, the solution in Laplace space has to fulfill the following condition,
\begin{align}
\left(\tilde{\rho}_S^{(u)}(\epsilon)\right)^\dagger = \tilde{\rho}_S^{(u)}(\epsilon^*) \; .
\end{align}
The solution (Eq.~(\ref{eq:solutionlaplace})) satisfies this under two constraints,
\begin{align}
1)& \quad \Sigma^\dagger(\epsilon) = \Sigma(\epsilon^*) \Rightarrow \Sigma^\dagger(\tau) = \Sigma(\tau) \label{eq:sigmacondition}\\
2)& \quad [\tilde{\rho}_S^{(u)}(\epsilon),-H_S+\Sigma(\epsilon)]=0 \; .
\end{align}
The first is fulfilled for the full self energy. The second constraint is achieved under the assumption that  $\tilde{\rho}_S^{(u)}(\tau=0) \propto \mathds{1}$.

To find the solution for the reduced density matrix, we have to perform the inverse Laplace transform,
\begin{align}
\tilde{\rho}_S^{(u)}(\tau) = \frac{1}{2\pi i}\int\limits_{\kappa-i\infty}^{\kappa+i\infty} \tilde{\rho}_S^{(u)}(\epsilon)\, e^{\epsilon\tau} \,\mathrm{d}\epsilon \; .\label{eq:inversetransformation}
\end{align}
This integral can be computed by using the residue theorem, where $\kappa$ lies in the region of convergence of $\tilde{\rho}_S^{(u)}(\epsilon)$.
We close the contour counter clockwise with a semicircle with infinite radius.
Therefore we enclose all singularities of $\tilde{\rho}_S^{(u)}(\epsilon)$.
The reduced density matrix can be written as a sum over all isolated singularities $i$,
\begin{align}
\tilde{\rho}_S^{(u)}(\tau) = \sum_i \Res_i{\left( e^{\tau\epsilon} \tilde{\rho}_S^{(u)}(\epsilon) \right)} \; .
\end{align}

To identify the singularities, we rewrite the solution for the reduced density matrix in Laplace space in terms of determinant and adjugate matrix,
\begin{align}
\tilde{\rho}^{(u)}_S(\epsilon) = \frac{1}{\det{(\epsilon+H_S-\Sigma(\epsilon))}}\adj{(\epsilon+H_S-\Sigma(\epsilon))} \; .
\end{align}
In the limit of small effect of the external degrees of freedom, hence small self energy, one can approximate the roots of the determinant as follows,
\begin{align}
\epsilon_s = - E_{s}+\Sigma_{ss}(- E_{s}) \; .
\end{align}
We introduce the renormalized energy $E_s^{re} = E_s-\Sigma_{ss}(-E_s)$.
It is also possible to find the roots of the determinant numerically and use the result as the renormalized energy.
For larger self energies with significant dependence on the energies $E_s$, it is possible that the determinant has more roots than the dimension of $H_S$.

The components of $\adj{(\epsilon+H_S-\Sigma(\epsilon))}$ are given as multiplications of components of $\epsilon+H_S-\Sigma(\epsilon)$.
Consequently the singularities of the adjugate matrix are the singularities of the self energy.
But this singularities coincide with the singularities of the determinant and are therefore canceled.
If we assume the remaining singularities to be simple poles, the reduced density matrix can be written as follows,
\begin{align}
\tilde{\rho}_S^{(u)}(\tau) = \sum_i e^{-\tau E^{re}_i} \frac{\adj{(-E^{re}_i+H_S-\Sigma(-E^{re}_i))}}{\prod_{j\neq i} (E^{re}_j - E^{re}_i) } \; . \label{eq:rhobetaresidue}
\end{align}
Compared to an equilibrium density matrix, the components of $\tilde{\rho}_S(\beta)$ retain the exponential factors but with a renormalized energy,
\begin{align}
\left(\tilde{\rho}_S(\beta)\right)_{ss'} = \sum_n f^{ss'}_n(\beta) e^{-\beta E_n^{re}} \; .
\end{align}
Furthermore the matrix is generally not diagonal in the eigenbasis of $H_S$.
For infinite coupling strength one expects $\tilde{\rho}_S(\beta)$ to commute with $H_C$\cite{Zurek1981,Review1982}. This transition of the eigenbasis is described by the values of $f^{ss'}_n(\beta)$ as a function of $\eta$.
To understand the influence of external degrees of freedom on specific model systems, these two effects can be studied by calculating the self energy and evaluating Eq.~(\ref{eq:rhobetaresidue}).

To illustrate the structure of the reduced density matrix and its dependence on the self energy $\Sigma$, we state the result for an arbitrary two dimensional system with non degenerate eigenenergies,
\begin{align}
&\tilde{\rho}^{(u)}_S(\beta) 
\propto e^{-\beta E_1^{re}} \nonumber\\ &\cdot\begin{pmatrix}
                                     E_2 -E_1^{re}-\Sigma_{22}(-E_1^{re}) & \Sigma_{21}(-E_1^{re})\\
                                     \Sigma_{12}(-E_1^{re}) & E_1-E_1^{re}-\Sigma_{11}(-E_1^{re})
                           \end{pmatrix}\nonumber\\
-& e^{-\beta E_2^{re}}     \nonumber\\  &\cdot\begin{pmatrix}
                                     E_2-E_2^{re}-\Sigma_{22}(-E_2^{re}) & \Sigma_{21}(-E_2^{re}) \\
                                     \Sigma_{12}(-E_2^{re}) & E_1-E_2^{re}-\Sigma_{11}(-E_2^{re})
                           \end{pmatrix} \; .
\label{eq:rhoin2d}
\end{align}
This equation shows how the renormalization of the energy and the change of the eigenbasis is expressed in terms of the full self energy.

\section{Lowest order results}\label{sec:lowest_order_results}
In this section we will present the results for the reduced density matrix with lowest order in the self energy.
For the spectral density of the bath modes, we assume an ohmic spectrum with Lorentzian cutoff at frequency $\omega_c$, whereas $\eta$ denotes the coupling strength (Eq.~\ref{eq:spectraldensity}).

In the standard master equation approach, lowest order results are normally obtained in the Markov approximation \cite{Karlewski2014}.
For an ohmic spectral density one can show that the bath correlator in imaginary time is not simply an exponentially decaying function, but has a peak at $\tau=0$ and $\tau=\beta$.
Therefore a Markov approximation is not suitable, and we calculate the self energy in Laplace space.
In lowest order we truncate the series for the self energy after the first term.
This term of the self energy is hermitian and therefore fulfills the condition (Eq.~(\ref{eq:sigmacondition})) for a hermitian density matrix.
The first order term of the self energy in Laplace space reads
\begin{align}
\Sigma^{(1)}(\epsilon) = \int\frac{\mathrm{d}\omega}{2\pi} \sum_{\bar{s}} \mathscr{O}\ket{\bar{s}}\bra{\bar{s}}\mathscr{O} \frac{S(\omega)}{E_{\bar{s}}+\epsilon+\omega} \; .
\end{align}
In the case of the ohmic spectral density with Lorentzian cutoff, the following integral has to be computed to calculate the self energy,
\begin{align}
I(E_s) = \int\limits_{-\infty}^{+\infty} \frac{d\omega}{2\pi} \frac{1}{E_s+\epsilon +\omega} \frac{\omega}{1-e^{-\beta\omega}} \frac{\eta}{1+(\omega/\omega_c)^2} \; .
\end{align}
This integral can be solved using the residue theorem.
For $\omega_c\neq \nu_n$ and $\epsilon \notin \{ -E_s-i \omega_c,-E_s-i \nu_n \}$ with $n\in\mathds{N}$, we find
\begin{align}
I(E_s) =& -i \frac{\epsilon+E_s}{1-e^{\beta(\epsilon+E_s)}} \frac{\eta}{1+\nicefrac{(\epsilon+E_s)^2}{\omega_c^2}} \Theta_{\nicefrac{1}{2}}(-\operatorname{Im}(\epsilon)) \nonumber\\ &+ i \frac{\omega_c^2 \eta}{2} \frac{1}{E_s+\epsilon+i\omega_c}\frac{1}{1-e^{-i\beta\omega_c}} \nonumber\\ &-\sum\limits_{n=1}^{\infty}\frac{\nu_n \eta}{\beta} \frac{1}{E_s+\epsilon+i\nu_n}\frac{1}{1-(\nicefrac{\nu_n}{\omega_c})^2} \; ,
\label{eq:integrallorenzian}
\end{align}
where we have introduced the Matsubara frequencies $\nu_n= \nicefrac{2\pi n}{\beta}$.
$\Theta_{\nicefrac{1}{2}}(x)$ describes the step function, which has the value $\nicefrac{1}{2}$ at $x=0$.
Thus we can construct $\Sigma^{(1)}(\epsilon)$ using the appropriate operator $\mathscr{O}$.
As described in the previous section, we can than calculate $\tilde{\rho}_S^{(u)}(\beta)$.

In the next subsections we apply these calculations to two qubit model systems. In the first model system we examine the influence of external degrees of freedom on an equilibrium expectation value as an example for a measurable quantity. With the second model system we show that even in thermal equilibrium external degrees of freedom can have an enormous influence on key quantum properties such as entanglement and coherence.
\subsection{Model system - 1 qubit}
Now we focus on a two-level system coupled to bosonic degrees of freedom,
\begin{align}
H_S=\frac{1}{2} \Delta E\,\sigma_z \quad H_B=\sum_i \omega_i b_i^\dagger b_i \; .
\end{align}
The coupling is characterized by the operator
\begin{align}
\mathscr{O}=\sigma_x+\sigma_z \; .
\end{align}

\begin{figure}[t]
\centering
\includegraphics{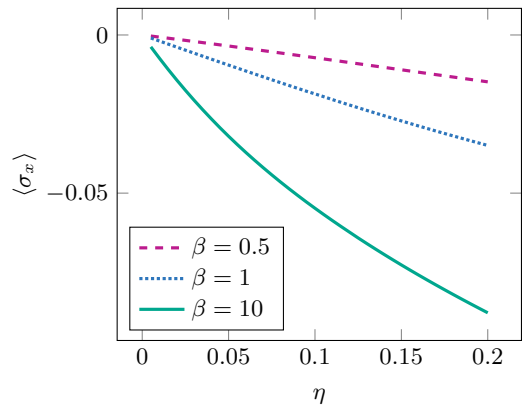}
\caption{This graph shows the expectation value of $\sigma_x$ as a function of the coupling strength $\eta$. Further parameters are $\Delta E=1$, $\omega_c=10$, $\beta=1$.}
\label{fig:expvaluesigmax}
\end{figure}

Using Eq.~(\ref{eq:integrallorenzian}) we find the first order self energy for the model system,
\begin{align}
\Sigma^{(1)}(\epsilon)=
                    \begin{pmatrix}
                    I(\nicefrac{-\Delta E}{2})+I(\nicefrac{\Delta E}{2}) & -I(\nicefrac{-\Delta E}{2})+I(\nicefrac{\Delta E}{2})\\
                   -I(\nicefrac{-\Delta E}{2})+I(\nicefrac{\Delta E}{2}) & I(\nicefrac{-\Delta E}{2})+I(\nicefrac{\Delta E}{2})
                    \end{pmatrix} \; .
\end{align}
With Eq.~(\ref{eq:rhoin2d}) we can now calculate the reduced density matrix.
Results for the expectation value of $\sigma_x$ are shown in Fig.~\ref{fig:expvaluesigmax}.
With increasing coupling strength $\braket{\sigma_x}$ decreases.
This illustrates, that the external degrees of freedom can have a substantial influence on the state of the system which can result in a significant change of measurable quantities.
%

\subsection{Model system - 6 qubits}
\begin{figure}[t]
\centering
\includegraphics{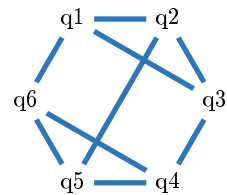}
\caption{The six qubit system is characterized by an Ising interaction for some qubit pairs. The interactions are indicated in this sketch. }
\label{fig:JijVerbindungen}
\end{figure}

We analyze a six qubit system similar to the eight qubit unit cell of a quantum annealing processor examined by Lanting et al. \cite{Lanting2014},
\begin{align}
H_S&=\epsilon \sum_{i<j} J_{ij}\, \xi_1^{(i,j)}\sigma_z^{(i)}\sigma_z^{(j)} -\frac{1}{2} \Delta \sum_i  \xi_2^{(i)} \sigma_x^{(i)} \; .
\end{align}
$J_{ij}$ is nonzero for the connections shown in Fig.~\ref{fig:JijVerbindungen}.
$\xi_{1/2}$ are Gaussian noise factors with mean value $1$.
We introduce this parameter noise to avoid degeneracy, because this would complicate the calculation of the residues compared to Eq.~(\ref{eq:rhobetaresidue}).
We assume all qubits to be coupled to individual bosonic baths ($b^{(i)}_k=b^{(j)}_k \, \forall i,j$),
\begin{align}
H_B&=\sum_{i,k} \omega_k \, b_k^{\dagger (i)} b^{(i)}_k \\
H_C&=\sum_{i,k}\sigma_z^{(i)} (b^{\dagger (i)}_k+b^{(i)}_k) \; .
\end{align}
Each bath has an ohmic spectral density with Lorentzian cutoff (Eq.~(\ref{eq:spectraldensity})).
As described in part~\ref{sec:generalization}, we can trace this problem back to the diagrammatic theory described for a simpler interaction Hamiltonian.

For the numerical simulation it is necessary to calculate the adjugate of a matrix. A potent improvement can be achieved by calculating the adjugate matrix using matrix decompositions\cite{Stewart1998}.

For the chosen set of parameters the ground state of $H_S$ is approximately the Greenberger-Horne-Zeilinger (GHZ) state\cite{Greenberger1990} of six qubits
\begin{align}
\ket{\psi_o} \approx \frac{1}{\sqrt{2}}\left(\ket{\uparrow\uparrow\dots} + \ket{\downarrow\downarrow\dots}\right) \; ,
\end{align}
which is a highly entangled state.
In the density matrix the entanglement of this state is expressed by matrix element $\braket{\uparrow\uparrow\dots|\rho_{GHZ}|\downarrow\downarrow\dots}$.

\begin{figure}[t]
\includegraphics{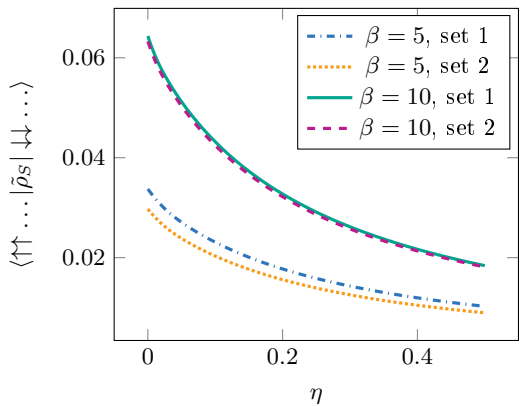}
\caption{The matrix element $\braket{\uparrow\uparrow\dots|\tilde{\rho}_S|\downarrow\downarrow\dots}$ decreases strongly for increasing coupling strength $\eta$. We show exemplary the results for two sets of random variables $\xi_1^{(i,j)}$, $\xi_2^{(i)}$ for each temperature. $\xi_{1/2}$ are Gaussian noise factors with variance $0.2$ / $0.1$. Further parameters are $\epsilon = 0.5$; $\Delta = 3$; $\omega_c = 15$; $J_{ij} = -3.01$ for the connections shown in the sketch, otherwise $J_{ij}=0$.}
\label{fig:Plot6Qubits}
\end{figure}
Fig.~\ref{fig:Plot6Qubits} shows this matrix element of the density matrix as a function of the coupling strength $\eta$.
For lower temperatures primarily the ground state is occupied.
Therefore $\braket{\uparrow\uparrow\dots|\tilde{\rho}_S|\downarrow\downarrow\dots}$ decreases for increasing temperatures because a state similar to $\frac{1}{\sqrt{2}}\left(\ket{\uparrow\uparrow\dots} - \ket{\downarrow\downarrow\dots}\right)$ also gets occupied.
For the one qubit system we examined the influence of the external degrees of freedom on an equilibrium expectation value.
Here we have focused on the variation of $\braket{\uparrow\uparrow\dots|\tilde{\rho}_S|\downarrow\downarrow\dots}$ for different $\eta$ to show that the coupling to a bosonic bath reduces the entanglement.
Each qubit is coupled to an individual bath via $\sigma_z^{(i)}$. For infinitely large coupling strength $\eta$ the density matrix of each qubit will be diagonal in the eigenbasis of $\sigma_z^{(i)}$.
Therefore we expect the density matrix of the six qubit system $\tilde{\rho}_S$ to approach a product state with increasing $\eta$.
This is reflected in the decay of the matrix element $\braket{\uparrow\uparrow\dots|\tilde{\rho}_S|\downarrow\downarrow\dots}$.
We see that even in thermal equilibrium a coupling to bosonic degrees of freedom leads to less quantum coherence.

\section{Conclusion}
Using a diagrammatic approach we derived an equation for the reduced density matrix of a system coupled to external degrees of freedom in thermal equilibrium.
It is possible to solve this equation in Laplace space.
The inverse Laplace transform leads to a renormalization of energies.
The renormalized energies can be approximated using the self energy or determined numerically.
Moreover we describe a change of the eigenbasis (Eq.~(\ref{eq:rhobetaresidue})) induced by the environmental degrees of freedom.
By analyzing a one qubit system we showed the fundamental influence of the bath on expectation values.
Furthermore, exemplified on a six qubit system, we worked out the impact of external degrees of freedom on key quantum properties such as entanglement and coherence in equilibrium.
For quantum emulation this means that external degrees of freedom can have a huge influence on the density matrix and therefore on the reliability of an emulator.


%

\end{document}

%% file: Abbildungen/tikzstyles.tex
\tikzset{
    particle/.style={
        postaction={decorate},
        decoration={
            markings,
            mark=at position .5 with {\arrow[xshift=2\pgflinewidth]{stealth reversed}}
        },
    },
}
\pgfkeys{
    selfen/.code 2 args={
        \node at (#1,#2) {\Large $\Sigma$};
        \draw[thick] (#1,#2) circle (0.25cm);
    }
}
\newcommand{\ww}[3]{ 
\draw[dashed] (#1,#3) arc[start angle=180, end angle=0, radius=(#2-#1)/2] -- (#2,#3);
}

%% file: Abbildungen/diagramme.tex
\draw[particle, very thick] (0.5,0) -- (1,0) node[xshift=0.25cm]{$=$};
\draw[particle] (1.5,0) -- (2,0) node[xshift=0.25cm]{$+$};
\draw[particle] (2.5,0) -- (3,0);
\draw[particle] (3,0) -- (3.5,0);
\draw[particle] (3.5,0) -- (4,0) node[xshift=0.25cm]{$+$};
\ww{3}{3.5}{0}
\draw[particle] (4.5,0) -- (5,0);
\draw[particle] (5,0) -- (5.5,0);
\draw[particle] (5.5,0) -- (6,0);
\draw[particle] (6,0) -- (6.5,0);
\draw[particle] (6.5,0) -- (7,0);
\ww{5}{5.5}{0}
\ww{6}{6.5}{0}
\draw[particle] ( 3,-1) node[xshift=-0.25cm]{$+$} -- (3.5,-1);
\draw[particle] (3.5,-1) -- ( 4,-1);
\draw[particle] ( 4,-1) -- (4.5,-1);
\draw[particle] (4.5,-1) -- ( 5,-1);
\draw[particle] ( 5,-1) -- (5.5,-1) node[xshift=0.25cm]{$+$}node[xshift=0.9cm]{$\dots \; .$};
\ww{3.5}{4.5}{-1}
\ww{4}{5}{-1}

%% file: Abbildungen/vollesrho.tex
\draw[particle, very thick] (0.5,0.5) -- (1,0.5);

%% file: Abbildungen/selbstenergie.tex
\pgfkeys{selfen={-0.25}{0}};
\node at (0.25,0) {$=$};
\draw[particle] (0.5,0) -- (1,0) node[xshift=0.25cm]{$+$};
\ww{0.5}{1}{0}
\draw[particle] (1.5,0) -- (2,0);
\draw[particle] (2,0) -- (2.5,0);
\draw[particle] (2.5,0) -- (3,0) node[xshift=0.25cm]{$+$} node[xshift=0.75cm]{\dots};
\ww{1.5}{2.5}{0}
\ww{2}{3}{0}

%% file: Abbildungen/dyson.tex
\draw[particle, very thick] (0.5,0) -- (1,0) node[xshift=0.25cm]{$=$};
\draw[particle] (1.5,0) -- (2,0) node[xshift=0.25cm]{$+$};
\draw[particle] (2.5,0) -- (3,0);
\pgfkeys{selfen={3.25}{0}};
\draw[particle, very thick] (3.5,0) -- (4,0);

%% file: Abbildungen/abbruch_reihe.tex
\draw[particle] (3.5,1.0) node[yshift=0.15cm,xshift=-0.25cm]{I:} -- (4.0,1.0);
\draw[particle] (4.0,1.0) -- (4.5,1.0);
\draw[particle] (4.5,1.0) -- (5.0,1.0);
\ww{4.0}{4.5}{1}
\draw[particle] (1.0,0.0) node[yshift=0.15cm,xshift=-0.25cm]{II:} -- (1.5,0.0);
\draw[particle] (1.5,0.0) -- (2.0,0.0);
\draw[particle] (2.0,0.0) -- (2.5,0.0);
\draw[particle] (2.5,0.0) -- (3.0,0.0);
\draw[particle] (2.5,0.0) -- (3.0,0.0);
\draw[particle] (3.0,0.0) -- (3.5,0.0)node[xshift=0.25cm]{,};
\ww{1.5}{2.5}{0}
\ww{2.0}{3.0}{0}
\draw[particle] (4.5,0.0) node[yshift=0.15cm,xshift=-0.25cm]{III:} -- (5.0,0.0);
\draw[particle] (5.0,0.0) -- (5.5,0.0);
\draw[particle] (5.5,0.0) -- (6.0,0.0);
\draw[particle] (6.0,0.0) -- (6.5,0.0);
\draw[particle] (6.5,0.0) -- (7.0,0.0);
\ww{5.0}{5.5}{0}
\ww{6.0}{6.5}{0}

%% file: Abbildungen/selbstenergie_abbruch.tex
\pgfkeys{selfen={-0.25}{0}};
\node at (0.25,0) {$=$};
\draw[particle] (0.5,0) -- node[below,yshift=-0.25em]{$\propto$ I} (1,0) node[xshift=0.25cm]{$+$};
\ww{0.5}{1}{0}
\draw[particle] (1.5,0) -- (2,0);
\draw[particle] (2,0) --node[below,yshift=-0.25em]{$\propto$ II} (2.5,0);
\draw[particle] (2.5,0) -- (3,0) node[xshift=0.25cm]{$+$};
\ww{1.5}{2.5}{0}
\ww{2}{3}{0}
\draw[particle] (3.5,0) -- (4,0);
\draw[particle] (4,0) --node[below,yshift=-0.25em]{$\propto$ II} (4.5,0);
\draw[particle] (4.5,0) -- (5,0) node[xshift=0.25cm]{$+$} node[xshift=0.75cm]{\dots};
\ww{3.5}{5}{0}
\ww{4}{4.5}{0}